\documentclass[10pt, conference, letterpaper]{ IEEEtran}
\IEEEoverridecommandlockouts

\usepackage{cite}
\usepackage{amsmath,amssymb,amsfonts}
\usepackage{algorithmic}
\usepackage{graphicx}
\usepackage{textcomp}
\usepackage{xcolor}
\usepackage{subfigure}
\usepackage{url} 
\usepackage{authblk}

\def\BibTeX{{\rm B\kern-.05em{\sc i\kern-.025em b}\kern-.08em
    T\kern-.1667em\lower.7ex\hbox{E}\kern-.125emX}}

\begin{document}
\pagestyle{plain}
\pagenumbering{arabic}
\title{\Large \bf HoneyGPT: Breaking the Trilemma 
in 
Honeypots with Large Language Models}

\author{
    {Ziyang Wang}\\
    Institute of Information Engineering, CAS\\
    School of Cyber Security, UCAS\\
    Beijing, China \\
    wangziyang2022@iie.ac.cn

    \and

    {Jianzhou You}\\
    Sangfor Technologies Inc.\\
    Shenzhen,China\\
    youjianzhou@gmail.com

    \and

    {Haining Wang}\\
    Virginia Tech Arlington\\
    Virginia, USA\\
    hnw@vt.edu

    \and

    {Tianwei Yuan}\\
    Institute of Information Engineering, CAS\\
    School of Cyber Security, UCAS\\
    Beijing, China \\
    yuantianwei@iie.ac.cn

    \and

    {Shichao Lv}\\
    Institute of Information Engineering, CAS\\
    School of Cyber Security, UCAS\\
    Beijing, China \\
    lvshichao@iie.ac.cn

    \and

    {Yang Wang}\\
    Shenzhen Institute of Advanced Technology,CAS\\
    Shenzhen, China\\
    yang.wang1@siat.ac.cn

    \and

    {Limin Sun}\\
    Institute of Information Engineering, CAS\\
    School of Cyber Security, UCAS\\
    Beijing, China \\
    sunlimin@iie.ac.cn
}

\maketitle

\thispagestyle{empty}


\begin{abstract}

Honeypots, as a strategic cyber-deception mechanism designed to emulate authentic interactions and bait unauthorized entities, often struggle with balancing flexibility, interaction depth, and deception. They typically fail to adapt to evolving attacker tactics, with limited engagement and information gathering. Fortunately, the emergent capabilities of large language models and innovative prompt-based engineering offer a transformative shift in honeypot technologies. This paper introduces HoneyGPT, a pioneering shell honeypot architecture based on ChatGPT, 
characterized by its cost-effectiveness and proactive engagement. In particular, we propose a structured prompt engineering framework that incorporates chain-of-thought tactics to improve long-term memory and robust security analytics, enhancing deception and engagement. Our evaluation of HoneyGPT comprises a baseline comparison based on a collected dataset and a three-month field evaluation. The baseline comparison demonstrates HoneyGPT's remarkable ability to strike a balance among flexibility, interaction depth, and deceptive capability. The field evaluation further validates HoneyGPT's superior performance in engaging attackers more deeply and capturing a wider array of novel attack vectors.
\end{abstract}

\section{Introduction}
Honeypots have been widely used as an effective approach for detecting, enticing, and understanding malicious activities on the Internet ~\cite{10.1145/3372297.3423356,kimequalnet,sasaki2022exposed,269257,wahab2019resource,rrushi2018dnic,torabi2020inferring,zhan2013characterizing,wang2023collaborative,huang2021duplicity}. 

These traps deceive attackers, enabling defenders to gather information, monitor behaviors for threat analysis, and prevent ongoing attacks. 
Terminal honeypots are a specialized type of honeypots for emulating real terminal systems to attract hackers or malicious software targeting terminal systems. Terminal honeypots range from low-interaction variants that simulate a minimal set of protocols or service information ~\cite{Honeyd}, to medium-interaction honeypots that replicate a broader array of system functionalities at a higher deployment and maintenance cost, and high-interaction honeypots that leverage sophisticated virtualization technologies, such as VMWare to create convincing environments of terminal systems or even deploy actual systems for maximum authenticity.

However, existing terminal honeypots face a trilemma problem related to flexibility, interaction level, and deceptive capability, stemming from the limitations of programmatic approaches and the fixed nature of system environments. In the trilemma, existing terminal honeypot designs struggle to achieve optimal performance across all three dimensions simultaneously, forcing defenders to compromise and select imperfect solutions.

It is very challenging to address the trilemma in practice. The primary methods for running honeypots are based on either programmatic simulations or real operating systems. On one hand, the programmatic simulation approach is limited by development costs, making it difficult to balance scalability and interaction depth. On the other hand, honeypots based on real operating systems suffer from rigidity, due to their inherent configurations that are short of flexibility. Moreover, 
both deployment methods lack intelligence during interactions, unable to provide responses that align with attackers' intentions to entice further engagement. This significantly limits the honeypot's deception capabilities. However, there has been a growing demand for honeypots that are more dynamic, intelligent, and cost-effective.

The development of Large Language Models (LLMs) \cite{kaplan2020scaling,wei2022emergent,YaoFu'sNotion,feng2024towards,NEURIPS2022_ebdb9904} appears to offer a potential solution to these problems. By providing attack commands as queries to the large language model, it can generate responses tailored to the attackers' interests based on the provided prompt. This approach breaks the trilemma of traditional honeypots, as a single LLM can simultaneously simulate multiple high-interaction honeypots with different configurations. Furthermore, by designing prompts to guide the LLM, it is possible to better fulfill attackers' objectives and entice them into deeper levels of interaction.

However, designing honeypots based on LLMs still faces three major challenges. 
A prominent issue is the absence of a standardized approach to honeypot prompt design. To address this problem, this paper establishes a universal specification for honeypot prompt keywords that adapts to long dialogues and enables real-time updates of dynamic memory during interactions. 
Another challenge arises from the native thinking constraints of LLMs, which cannot handle complex attack combinations. We employ a question enhancement strategy based on CoT that decomposes the honeypot's long-dialogue response tasks into three sub-tasks, allowing LLMs to assess the command's impact on the operating system during each interaction. When posing the next question, we integrate the consolidated changes in the operating system and the user command when posing the next question to improve its understanding of complex attack combinations.
The third challenge lies in that the performance of LLMs in prolonged sessions is hampered by their context length limitations and inherent forgetting mechanisms. To address this issue, we employ a Memory Pruning algorithm based on CoT that strategically prunes the session history based on its significance and temporal characteristics, ensuring that only the most essential contextual information is retained within the confines of the context length, thereby optimizing the honeypot's effectiveness in extended interactions. The assessment of each attacker's significance is performed by LLMs to support the automatic operation of the honeypot.


The main contributions of this work are summarized below.

1. We propose HoneyGPT, an intelligent terminal honeypot that addresses the dimensions of flexibility, interaction level, and deceptive capability, enabling extended interactions while overcoming the trilemma.

2. We propose a question enhancement strategy and a Memory Pruning algorithm based on the Chain of Thought (CoT) approach for terminal honeypot task scenarios. These techniques enhance the task-solving capabilities of native LLMs and balance the honeypot's effectiveness and conversational consistency within the inherent context length limitations in extended interactions. The introduction of these strategies is essential to advancing honeypot technology and improving network security defense capabilities.

3. We conduct a baseline evaluation of HoneyGPT in comparison with traditional honeypots, utilizing open-source attack data captured by Cowrie. The results indicate that HoneyGPT outperforms conventional honeypots with respect to its composite capabilities in flexibility, level of interaction, and deception efficacy.

4. We deploy HoneyGPT and Cowire, which is a programmatic honeypot, on the  Internet for three months
 and observe that HoneyGPT captures more attack behaviors and longer interactions than Cowrie. 

The rest of this paper is structured as follows. Section 2 provides background on LLMs, prompt engineering, and terminal honeypots. Section 3 describes the trilemma of flexibility, interaction, and deception in existing terminal honeypots. Section 4 presents our HoneyGPT solution, detailing its prompt design, strategy, workflow, and configuration. 
Section 5 validates the efficacy of HoneyGPT, providing baseline comparisons on deception, interaction, and flexibility, as well as field evaluations to measure its effectiveness in engaging attackers and capturing novel attack vectors. Section 6 outlines the limitations and potential future work, and finally, Section 7 concludes the paper.

\section{Background}
In this section, we provide the background 
of Large Language Models, prompt engineering, and terminal honeypots.

\subsection{Large Language Models}
Large Language Models (LLMs) are a subset of pre-trained language models (PLMs), characterized by their voluminous parameter sets, reaching or even exceeding hundreds of billions, as defined by Kaplan et al \cite{kaplan2020scaling}. Since OpenAI introduced  ChatGPT ~\cite{openai.com}, 
LLMs have been widely used across a spectrum of disciplines ~\cite{YaoFu'sNotion,brown2020language}, particularly in the field of cybersecurity ~\cite{meng2024large,hudegpt,he2023large,pearce2023examining,pearce2022asleep,vishwamitra2024moderating}. Compared to previous models, LLMs offer two main advantages in interdisciplinary applications: 

Firstly, LLMs display astonishing emergent abilities as identified by Wei et al \cite{wei2022emergent}, which are competencies absent in smaller-scale models but manifest in larger configurations. Such models experience a quantum leap in performance relative to their diminutive counterparts. Prominent among these emergent abilities are:
\textbf{In-context learning}: LLMs can, upon receiving natural language instructions or several task demonstrations, generate accurate outputs for test cases by continuing the input text sequence, circumventing the need for further training or gradient adjustments ~\cite{brown2020language}.
\textbf{Instruction following}: LLMs possess the ability to comprehend and execute natural language instructions with minimal or even zero examples, adapting to new tasks~\cite{victor2022multitask}. By fine-tuning tasks articulated through instructional prompts, LLMs have demonstrated proficiency in accurately handling tasks they have never seen before~\cite{ouyang2022training}.
\textbf{Step-by-step reasoning}: Distinct from conventional PLMs, LLMs are capable of deconstructing complex tasks into a sequence of discrete subtasks by employing prompt-based strategies such as chain-of-thought. It is conjectured that this capacity may be derived from their training on code ~\cite{wei2022chain,YaoFu'sNotion}.

Secondly, LLMs transform the traditional employment of AI algorithm development and utilization, significantly diminishing the technical threshold for cross-disciplinary actors to harness such algorithms. In contrast to previous models, users engage with LLMs via a prompt interface (e.g., the GPT-3-turbo API), by formulating natural language prompts or directives to modulate the model's behavior to yield anticipated results. This empowers individuals who lack a deep understanding of model training paradigms but are conversant with prompt engineering to employ LLMs for task facilitation.

\subsection{Prompt Engineering}
Prompt engineering is the process of crafting prompts to better guide large language models to understand and resolve tasks. Here we briefly introduce prompt engineering from its core facets: components, design principles, and strategic approaches.

\textbf{Components}.
The standard components of LLM prompts typically include:
1. Task Description: Instructions that the LLM is expected to adhere to, expressed in natural language.
2. Input Data: The requisite data for a task, presented in natural language.
3. Contextual Information: The contextual or background information that can aid in the execution of a task.
4. Prompt Style: The formulation of the prompt, which is tailored to optimize the model's responses, such as through role-playing or incremental reasoning.

\textbf{Design Principles}.
The development of effective LLM prompts is predicated on:
1. Clarity: Ensure instructions are explicit and communication is clear through symbolic delimiters, structured outputs, conditional logic, and "Few-shot" examples.
2. Deliberation: 
Allow LLMs time to reason by structuring prompts for sequential information processing using "chain of thought" techniques.

\textbf{Prompting Strategies}.
The "Chain of Thought" (CoT) ~\cite{wei2022chain,feng2024towards,NEURIPS2022_ebdb9904,ling2024deductive,lu2022learn,yang2022generating,zelikman2022star} paradigm has been shown to significantly enhance LLM performance on tasks that require complex reasoning by incorporating intermediary reasoning steps within the prompts. The zero-shot CoT, as introduced in foundational research, employs prompts such as "Let's think step by step" to shepherd LLMs to a logical conclusion, thus highlighting the significance of scale in emergent capabilities.

\subsection{Terminal Honeypots}
Honeypots are security tools designed as decoys to attract and monitor unauthorized or malicious attackers. 
Based on their implementation approaches, terminal honeypots can be categorized into three primary types: emulated honeypots, real-system honeypots, and hybrid honeypots.

\textbf{Emulated Honeypots}
are software constructs that simulate network services and devices to attract cyber adversaries using standard programming techniques. 
These honeypots range from low (Honeyd~\cite{Honeyd}) to medium (Cowrie\cite{Cowrie}) interactivity but face constraints due to high development and operational costs, with trade-offs in adaptability and increased costs. Such frameworks have laid the groundwork for a range of specialized offshoots~\cite{10.1145/3372297.3423356,220564}.

\textbf{Real-system Honeypots}    
utilize genuine, unmodified system environments to provide a high level of interactions~\cite{wang2022camshield,269257,wahab2019resource}. This approach enhances the detection of complex attacks but suffers from high deployment and maintenance costs and reduced adaptability due to their rigidity~\cite{220564}.

\textbf{Hybrid Honeypots}  
aim to merge the advantages of emulated and real-system honeypots through a scalable hybrid framework~\cite{xu2021brief,you2021honeyvp}.  This framework employs programmed methods for simulating low-level interactions, while delegating high-level interactions to real file system. While this enhances flexibility and interactivity, it also  faces challenges in harmonizing simulated protocols with actual system responses, impacting their effectiveness against sophisticated attacks.

\section{Trilemma of Existing Terminal Honeypots}

The Trilemma of Terminal Honeypots reflects the critical balance required between interaction depth, flexibility, and deception capability. 
Honeypots from their inception have grappled with this trilemma, consistently facing trade-offs that impede their overall efficacy.

\textbf{Limited Flexibility}
refers to the system's adaptability in authentically replicating a variety of systems and services, which is pivotal but challenging. Emulated honeypots incur high costs and struggle with seamlessly transitioning across varied terminal environments while maintaining high interaction fidelity. Real-system honeypots lack the necessary adaptability and carry significant management overhead. The main challenge lies in designing honeypots that balance realistic interactions and adaptability without excessive costs or management burdens.

\textbf{Limited Interaction} 
is a result of limitations in implementation methods and development costs. Emulated honeypots are limited to simple rule matching and cannot execute complex command sequences like a real OS, reducing their interactive capability and effectiveness in engaging sophisticated attackers. Enhancing honeypot interaction is essential for improving their reliability and effectiveness.

\textbf{Limited Deception} is often due to the static and inflexible behaviors of existing honeypot systems. Emulated honeypots, restricted by static configurations and predefined rules, fail to simulate complex interactions, making them easily identifiable to experienced attackers. Similarly, real-system honeypots, despite using authentic environments, are constrained by cost and rigidity, offering limited configurations and restricted permissions that do not sufficiently engage attackers, and thus failing to entice attackers into a more profound engagement.

\section{HoneyGPT}
To address the trilemma of existing honeypots, we present HoneyGPT, an advanced honeypot framework based on ChatGPT.
HoneyGPT transforms the conventional "request-response" message interaction (based on terminal protocols) into a "question-answer" text interaction (via ChatGPT API). 

\begin{figure}
\vspace{-0.1in}
    \centering
    \includegraphics[width=1\linewidth]{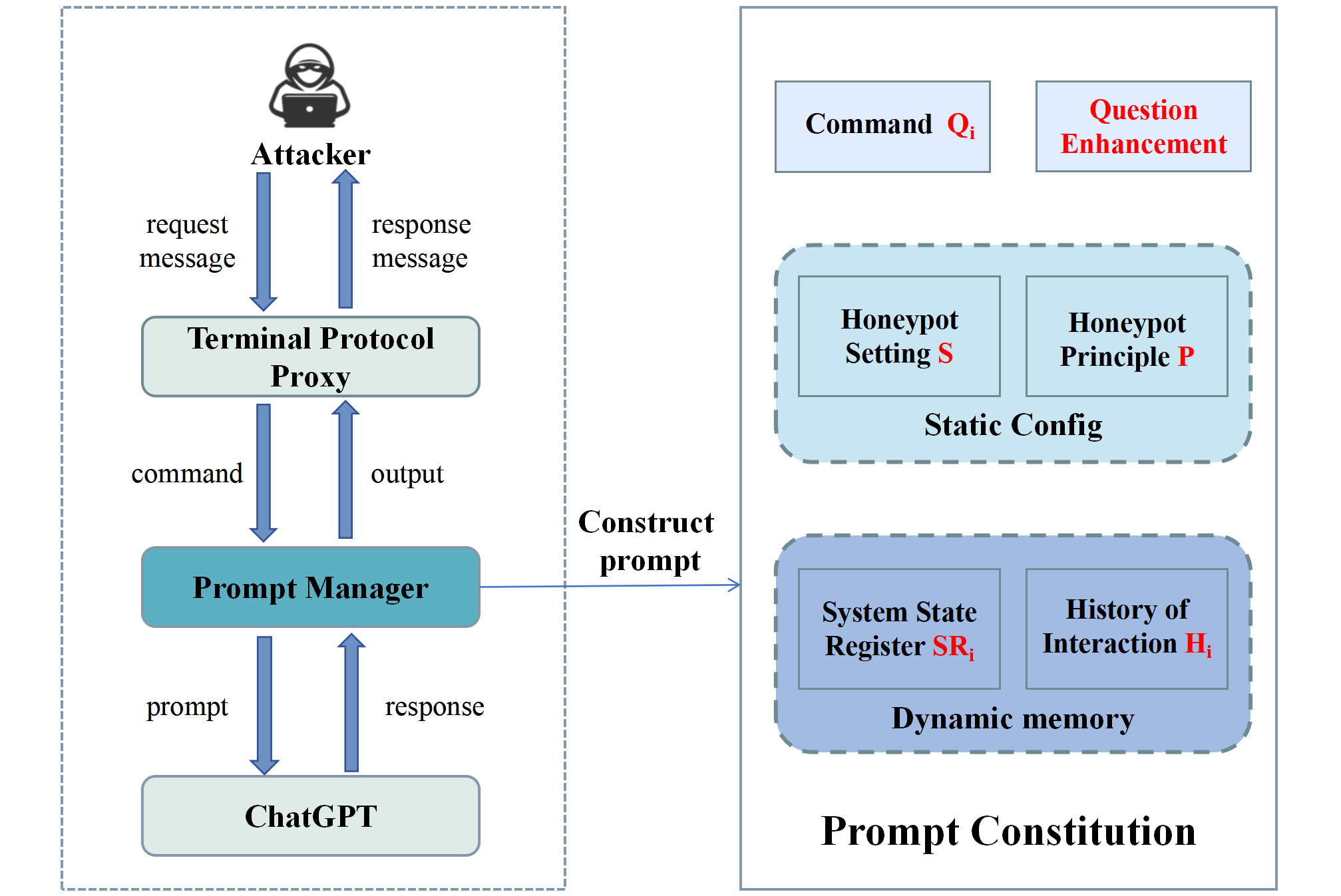}
    \vspace{-0.3in}
    \caption{HoneyGPT Framework}
    \label{fig:HoneyGPT-framework}
\vspace{-0.2in}
\end{figure}

As shown in Figure~\ref{fig:HoneyGPT-framework}, the framework consists of three main components: the Terminal Protocol Proxy, the Prompt Manager, and ChatGPT. The Terminal Protocol Proxy parses attackers' commands from request messages and forwards them to the Prompt Manager. Additionally, it encapsulates terminal output received from the Prompt Manager into response messages, which are then sent back to the attackers. 
The Prompt Manager generates prompts to interact with ChatGPT and, based on the response, extracts the terminal output to be returned to the attackers while maintaining the state of honeypot. It employs CoT-based question enhancement and memory pruning mechanisms to improve the consistency and authenticity of the honeypot during ongoing interactions.
ChatGPT generates response outputs based on the provided prompts, which are then analyzed by the Prompt Manager for further processing and state management. Since the Terminal Protocol Proxy just repurposes the work of the Cowrie honeypot, we mainly elaborate on the design of the Prompt Manager in the following.

\begin{figure*}
    \centering
    \includegraphics[width=.8\linewidth]{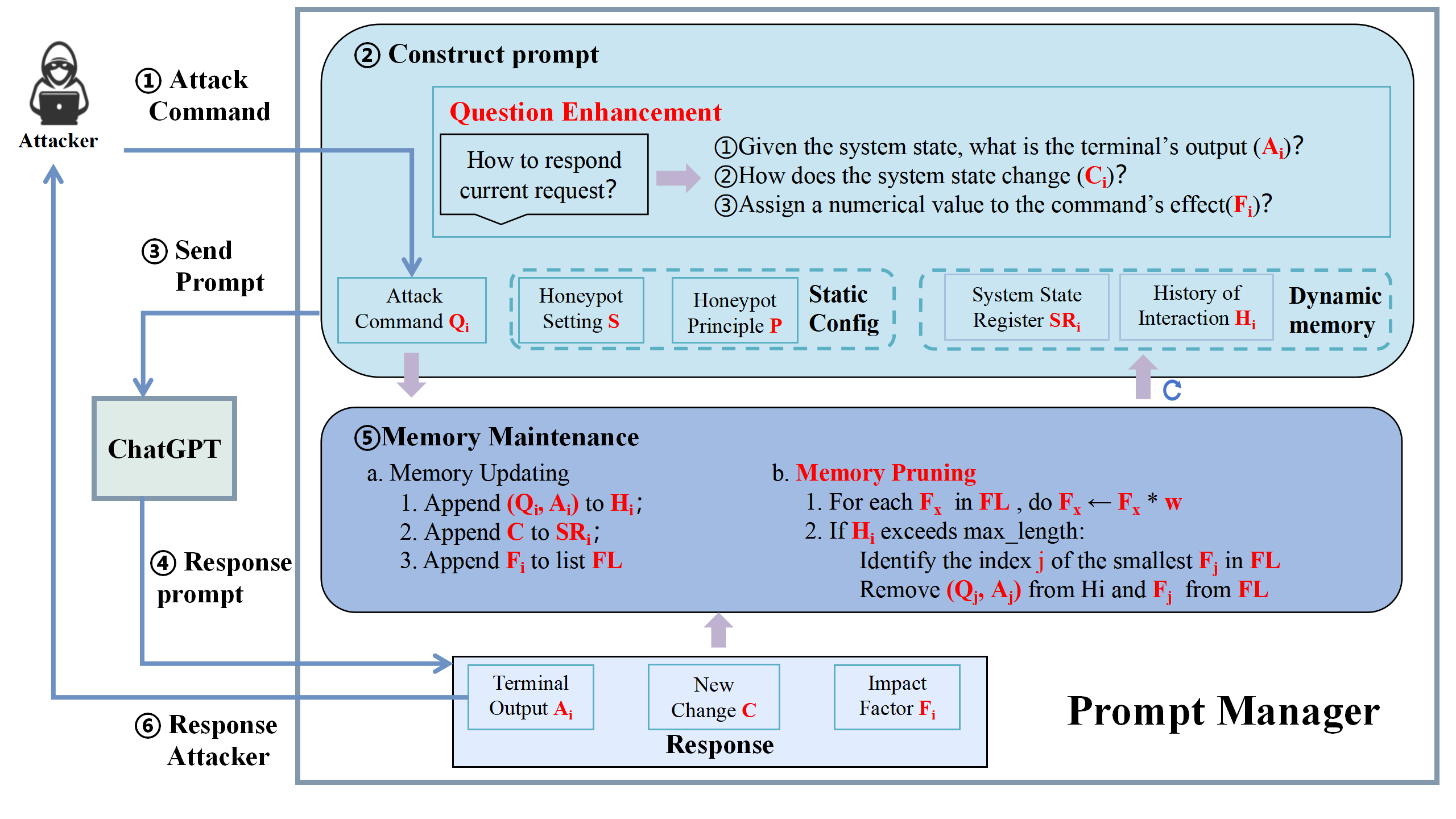}
    \vspace{-0.2in}
    \caption{HoneyGPT workflow}
    \label{fig:HoneyGPT workflow} 
\vspace{-0.2in}
\end{figure*}

\subsection{The Constitution of Prompt} \label {The constitution of prompt}
We first  define ChatGPT's interaction history as a sequence of N question-answer pairs: 
\begin{equation}\{(Q_1, A_1),(Q_2, A_2), \ldots, (Q_{N}, A_{N})\}\end{equation}

To obtain the response $A_i$ from the $i$-th interaction, we construct complex prompts. The $i$-th HoneyGPT interaction is formally defined by Equation ~\ref{eq:ChatGPT}.
\begin{equation} \label{eq:ChatGPT} 
(A_i, C_i, F_i) = ChatGPT(P, S, Q_i, SR_i, H_i)
\end{equation}

As shown in Figure~\ref{fig:HoneyGPT-framework}, the constitution of the prompt for the $i$-th interaction encompasses six components: Command \(Q_i\) , Question Enhancement, Honeypot Principle \(P\), Honeypot Setting \(S\), System State Register \(SR_i\) , and History of Interaction \(H_i\). Here \(P\) and \(S\) represent the static configuration of the prompt, remaining unchanged throughout the interactions. \(SR_i\) and \(H_i\) are dynamic memory aspects of the prompt, which are updated after each interaction by the Prompt Manager. 
The definitions of variables used in HoneyGPT are given below.
\begin{itemize}
\item \textbf{Honeypot Principle \(P\)} 
defines the Principle that the LLMs should follow when performing honeypot tasks

\item \textbf{Honeypot Setting \(S\)} 
refers to the Honeypot initial configuration set.
\item \textbf{Attacker Query \(Q_i\)} 
denotes the command issued by the attacker during the $i$-th interaction.

\item \textbf{Honeypot Answer \(A_i\)} 
denotes the terminal output returned to the attacker during the $i$-th interaction.

\item \textbf{History of Interaction \(H_i\)}, represents the record of interactions between attackers and the honeypot prior to the i-th interaction, is defined as follows:
\begin{math}
\{(Q_1, A_1), (Q_2, A_2), \ldots, (Q_{i-1}, A_{i-1})\}
\end{math}. 

\item \textbf{System State Register \(SR_i\)} 
 refers to a collection of system states communicated to LLMs at the $i$-th interaction:
 \begin{math}
    \{(C_1), (C_2), \ldots, (C_{i-1})\},
 \end{math}
where each state \(C_i\) corresponds to the system's changes after each prior interaction, paralleling the order of \(H_i\). 

\item \textbf{New Change \(C_i\)} denotes the changes to the system environment after $i$-th interaction.

\item \textbf{Impact Factor \(F_i\)} quantifies the impact of \(Q_i\) on the system.

\item \textbf{Weaken Factor \(w\)}, utilized in the pruning algorithm, systematically diminishes the efficacy of \(Q_i\) by scaling it with \(w\) after each interaction.
\end{itemize}

\subsection{The Strategy of Prompt Manager}

The design of HoneyGPT introduces strategies based on the Chain of Thought (CoT) for question enhancement and memory pruning. These strategies are developed to address the intrinsic limitations of ChatGPT in processing complex, extended dialogues and to circumvent the constraints of the prompt context length. The essence of CoT lies in compelling the LLM to decompose a multifaceted problem into a sequence of sub-problems, tackling them incrementally to enhance the LLM’s ability to comprehend and address intricate issues.

\subsubsection{\textbf{Question Enhancement for Complex Extended Dialogue Tasks}} \label{Question Enhancement}
\begin{figure}[t]
    \centering
    \includegraphics[width=1.0\linewidth]{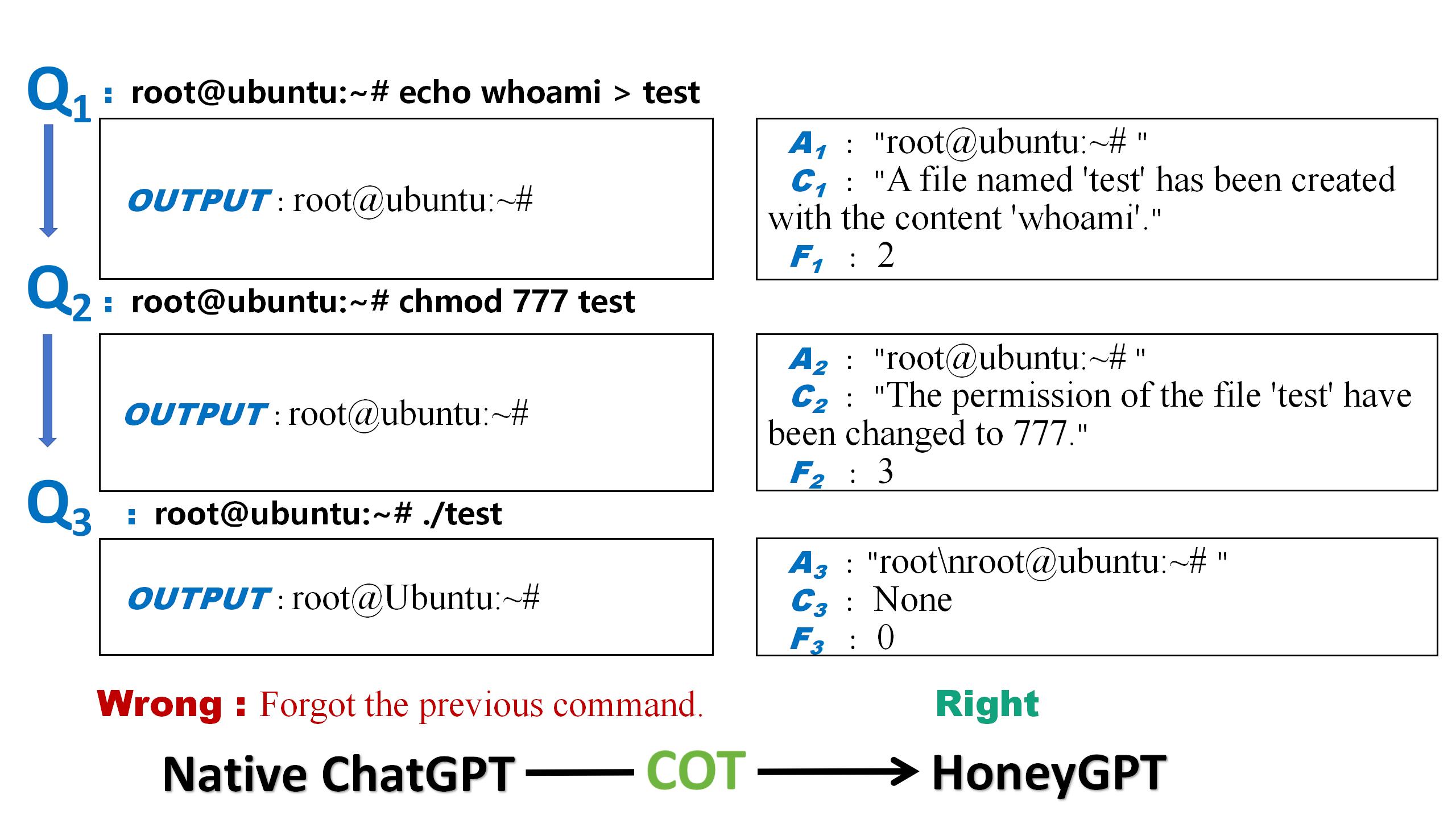}
    \vspace{-0.3in}
    \caption{Comparison of the Effects of Question Enhancement}
    \label{fig:COT}
    \vspace{-0.1in}
\end{figure}

Adversarial scenarios often involve attackers executing a series of malicious commands, such as the "write-elevate-execute" sequence depicted in Figure \ref{fig:COT}. Native ChatGPT, while adept with individual commands, struggles with complex, interdependent dialogues requiring coherence and memory. In contrast, HoneyGPT's integration of a Question Enhancement strategy markedly improves its capacity to process these complex sequences. Within HoneyGPT, we enhance the rudimentary long-dialogue response tasks of the honeypot into three specific sub-tasks:
\begin{itemize} 
\item Given the system state, what is the terminal’s output ($A_i$) ?

\item How does the system state change ($C_i$) ?

\item Assign a numerical value to the command’s effect ($F_i$).
\end{itemize}

To ensure the continuity and automation of the attack interaction process, the tasks of evaluating the impact of attack commands on the operating system, namely questions 2 and 3, have been delegated to ChatGPT. The analysis results from these tasks provide support for question 1 by supplying system state information in subsequent sessions. The question enhancement mechanism refines the questions and leverages ChatGPT to analyze system changes, thereby enhancing its consistency and effectiveness in handling complex, long-dialogue issues.


\subsubsection{\textbf{Memory Pruning for Prompt Length Limitation}}
As interactions accumulate, the dynamic memory ($H_i$) increases, leading to a longer prompt for subsequent interactions. 
However, due to limitations imposed by computational resources, storage capacities, and the inherent characteristics of the self-attention mechanism, LLMs have a restricted context length for prompts. Exceeding this limit renders ChatGPT non-functional. To address this issue, we introduce Memory Pruning algorithm: when the context length of a prompt surpasses the predefined threshold, we prune the least impactful entries from $H_i$ to balance the consistency and usability of HoneyGPT. While the $SR_i$ also increases dynamically, its length remains well below the context length limit, thus obviating the need for its pruning. The impact of $H_i$'s each entry is assessed by ChatGPT using the quantitative criteria specified in Table \ref{tab:Rules of the Assignment of Values to $F_i$}. 
Furthermore, to account for the diminishing influence of earlier commands on the Impact Factor ($F_i$) over time, we introduce a Weaken Factor ($w$) that is applied to each interaction's $F_i$. When selecting the value of $w$, it is crucial to ensure that a newly executed command carries more weight than a high-privilege command that occurred three steps prior. This principle can be represented mathematically as \( F_{i+3} \cdot w^3 > F_i \). 
Moreover, the value of $w$ should not be excessively small, as it must adequately capture the effect of time on the relevance of past commands.

\begin{table}[t]
    \caption{Rules of the Assignment of Values to $F_i$}
    \vspace{-0.1in}
    \label{tab:Rules of the Assignment of Values to $F_i$}
    \begin{tabular}{p{0.5cm}p{7cm}}
        \hline
        Value & Condition\\
        \hline
        0 & Read file, display system information\\
        1 & Create file, install tool\\
        2 & Modify files/dir, change working directory, change shell\\
        3 & Start/stop service, download file, Elevate privilige\\
        4 & Impact services, delete files, password changed\\
        \hline
    \end{tabular}
    \vspace{-0.2in}
\end{table}

\subsection{The Workflow of Prompt Manager}
As shown in Figure \ref{fig:HoneyGPT workflow}, The Prompt Manager orchestrates the interaction cycle in HoneyGPT, systematically handling attack commands and constructing precise prompts. This section delineates the workflow, emphasizing the integration of dynamic memory management and question enhancement strategies. Given that the Terminal Protocol Proxy solely handles protocol parsing and encapsulation, its discussion is omitted here. The workflow of the Prompt Manager is as follows:
\textbf{(1) Receive Attack Command:} The attacker sends an attack command, denoted as $Q_i$.
\textbf{(2) Construct Prompt:} The Prompt Manager constructs a comprehensive prompt, integrating the Question Enhancement mechanism~\ref{Question Enhancement}, which decomposes the interaction task into three sub-tasks.Additionally, the prompt includes the attacker's command \(Q_i\), and combines static configuration elements such as \(S\) and \(P\) with dynamic memory components \(SR_i\) and \(H_i\).
\textbf{(3) Interaction with ChatGPT:} The constructed prompt is dispatched to ChatGPT for processing.
\textbf{(4) Receive Response from ChatGPT:} ChatGPT returns a response, which includes the terminal output \(A_i\), system changes \(C_i\), and impact factor \(F_i\).
\textbf{(5) Memory Maintenance:} is divided into Memory Updating and Memory Pruning. Memory Updating integrates the current session's memory into the prompt's dynamic memory. Specifically, it appends \((Q_i, A_i)\) to \(H_i\) and adds \(C_i\) to \(SR_i\). Additionally, \(F_i\) is stored in the list \(FL\) to inform Memory Pruning. Memory Pruning manages the prompt's length. It first applies time decay by multiplying each session record's impact factor in \(FL\) by the Weaken Factor \(w\). Next, it checks if the prompt length is approaching its limit. If it is, it identifies the index \(j\) of the smallest \(F_j\) in \(FL\) and removes \((Q_j, A_j)\) from \(H_i\) and \(F_j\) from \(FL\) to maintain efficiency.
\textbf{(6) Response attacker:} The refined response \(A_i\) is delivered back to the attacker, concluding the interaction sequence with a well-informed, precise output.

\subsection{Static Configuration of HoneyGPT}
Here we introduce the static components involved in configuring HoneyGPT, namely the System Principles (\(P\)) and Honeypot Settings (\(S\)), which remain unchanged once a honeypot is initialized.

\subsubsection{System Principles $P$}
System Principles provide behavioral guidelines for simulating system activities, covering aspects such as the role of HoneyGPT, time sensitivity, input/output formats, and few-shot learning. These principles enable HoneyGPT to effectively attract and analyze potential attack behaviors by maintaining a credible system illusion. HoneyGPT, serving as a terminal honeypot system, is specifically designed to attract and monitor attack activities. To achieve this objective, we have implemented the system's principles within the prompts, providing operational guidelines for ChatGPT. These prompts play different crucial roles in the functionality of the system, including:

\textbf{ Role of HoneyGPT}
HoneyGPT strategically simulates an authentic terminal environment, tailored to align closely with attackers' intents, thereby encouraging deeper and more sustained engagement. The system expertly balances authenticity with the objective of enticing attackers into prolonged interactions, thereby enhancing the honeypot's effectiveness.

\textbf{ Time Sensitivity}
In scenarios where attacks are time-sensitive, attackers may issue commands related to current time inquiries, such as uptime and top. Since ChatGPT inherently lacks the capability to query real-time networked time data, it is necessary for us to provide time-related information such as the current time and boot time in the prompts to ChatGPT. This approach ensures that the honeypot's responses are temporally accurate, thus more convincingly mimicking a real system environment.

\textbf{ Input/Output Format} 
The structure of both input and output prompts critically defines each interaction with the LLM in HoneyGPT. The input format is crucial for the LLM's accurate understanding of the honeypot system's state and the commands issued by attackers. Conversely, the output format affects HoneyGPT's operational effectiveness. To ensure HoneyGPT's robustness, we instruct the LLM to use the JSON format for outputs. This standardization ensures consistent and correct parsing of responses during each interaction with the model.

\textbf{ Few-Shot Learning} 
HoneyGPT utilizes a few-shot learning strategy to enhance ChatGPT's comprehension of task requirements and improve the quality of its outputs, compensating for gaps in several command knowledge. Our 
experience indicates that by integrating only 4-5 representative examples, HoneyGPT's generative capabilities can be significantly enhanced. 

\subsubsection{Honeypot Setting $S$}
The Honeypot Settings describe the terminal system information simulated by HoneyGPT, which is crucial for attracting and capturing potential attackers, involving specific configurations like hardware specifications and software environments. These settings are designed to enhance the honeypot's attractiveness and precision in capturing malicious exploitation such as cryptocurrency mining. 
Our experience indicates that honeypots with high-end configurations are more attractive to malicious entities, such as cryptocurrency-mining malware. Furthermore, the more details of honeypot configurations, the higher precision of the responses. The ability to customize honeypot settings via natural language significantly enhances flexibility and reduces the complexity of deployment. Honeypot settings are divided into two main categories: 
hardware and software.

\textbf{ Hardware} 
It is essential to define the hardware specifications that will be emulated by HoneyGPT. It includes defining CPU types and counts, GPU presence, and storage capacities, those attributes that attackers frequently scrutinize. The specifications selected must mirror those of the systems intended to be simulated by HoneyGPT to maintain the authenticity of the honeypot environment. Additionally, configuring systems with high-end GPUs and CPUs can attract specific types of threat actors, such as crypto-mining malware, prompting deeper interactions.

\textbf{ Software} 
The software environment of a honeypot, including the operating system, open ports, user configurations, and application services, must be strategically crafted to reflect the behaviors of potential targets. For instance, if the objective is to mimic a web server, HoneyGPT should be configured with the appropriate web server software, along with associated services and open ports. Additionally, attackers are often drawn to details such as system processes, resource utilization, scheduled tasks, user configurations, and filesystem information. These elements can be specifically tailored within HoneyGPT to create customized responses for each engagement scenario.

\section{Evaluation}
Our evaluation of HoneyGPT includes two parts: the baseline evaluation and the field evaluation. The baseline evaluation compares the capabilities in deception, interaction level, and flexibility between HoneyGPT and traditional honeypots. Specifically, to evaluate the honeypots' capability in deception and interaction level under consistent attack scenarios, the baseline evaluation utilizes the attack dataset collected by Cowrie. The field evaluation involves deploying HoneyGPT and Cowrie simultaneously on the Internet for a period of three months to assess whether HoneyGPT can influence attackers' attack strategies and capture a wider range of attack types and interaction lengths.

\subsection{Baseline Comparison Set}
\begin{table}[t]
\vspace{-0.6cm}
    \caption{Baseline Comparison Set}
    \label{tab:Baseline comparison  Set}
    \begin{tabular}{p{1.2cm}p{3cm}p{1.4cm}p{1.3cm}}
    \hline
Capability Type & Subjects & Method & Research Type  \\
\hline
Deception          & HoneyGPT (GPT-3.5 turbo), HoneyGPT (GPT-4),  Cowrie, real-system                      & Replay attack dataset  & Quantitative Research \\
\hline
Interaction  & HoneyGPT (GPT-3.5 turbo), HoneyGPT (GPT-4),  Cowrie& Replay attack dataset  & Quantitative Research \\
\hline
Flexibility        & HoneyGPT, Cowrie, Honeyd, real-system                                                & Verification Individually  & Qualitative Research  \\
\hline
\end{tabular}
\vspace{-0.2cm}
\end{table}

The comparison in the baseline evaluation is from three aspects: deception, interaction level, and flexibility. The comparisons of deception and interaction level are performed in a quantitative manner, by utilizing the same attack dataset to ensure the same attack scenarios, while the flexibility comparison is performed in a 
qualitative manner, through individual experimental verification for each indicator.

Table \ref{tab:Baseline comparison Set} shows the experimental setup for the baseline comparison. The comparison of deception is conducted among HoneyGPT based on GPT-3.5 turbo interface, HoneyGPT based on GPT-4 turbo interface, Cowrie, and a real system. The interaction level comparison is limited to the comparisons among HoneyGPT based on GPT-3.5 turbo interface, HoneyGPT based on GPT-4 turbo interface, and Cowrie, because the real system exhibits the best interaction level and there is no need to compare. 
The flexibility comparison is conducted among Honeyd, a low-interaction honeypot, along with HoneyGPT, Cowrie, and the real system.

\begin{figure*}[t]
\vspace{-0.1in}
	\centering
\subfigure[SALC,SALNLC,FALC,FALNLC\label{fig:SALC,SALNLC,FALC,FALNLC}] 
    {\includegraphics[width=.4\textwidth]{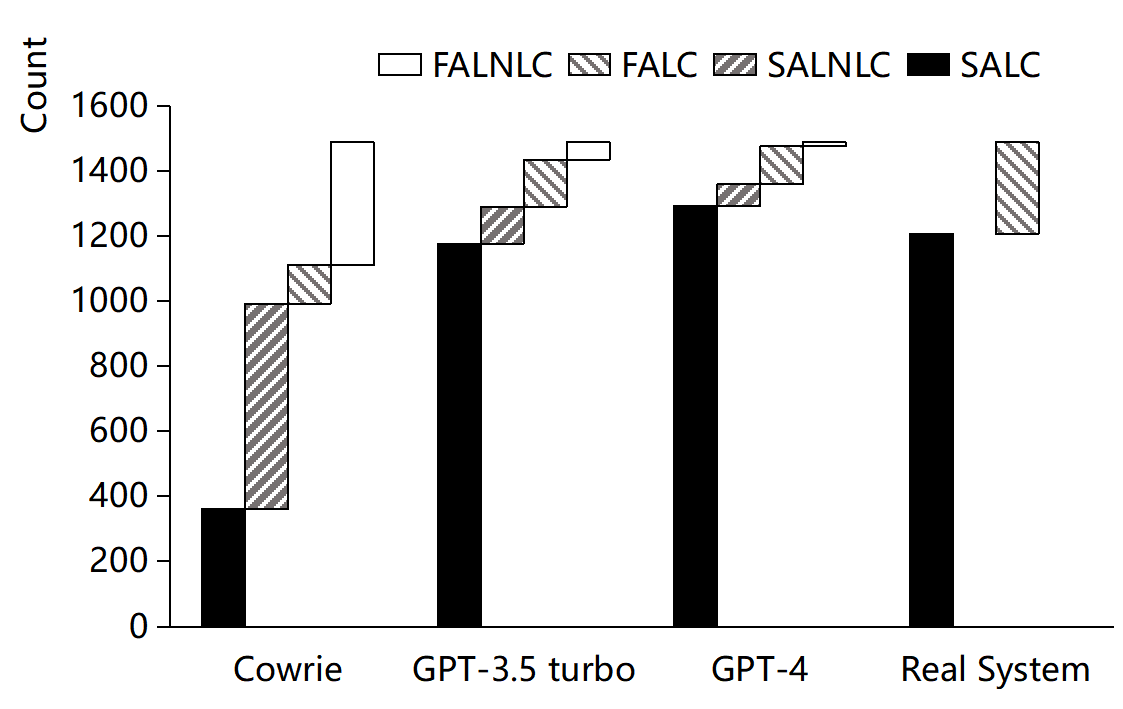}}
\subfigure[Accuracy,Temptation,Attack Success Rate,OS Logic Compliance\label{fig:Accuracy,Temptation,Attack Success Rate,OS Logic Compliance}]      {\includegraphics[width=.4\textwidth]              {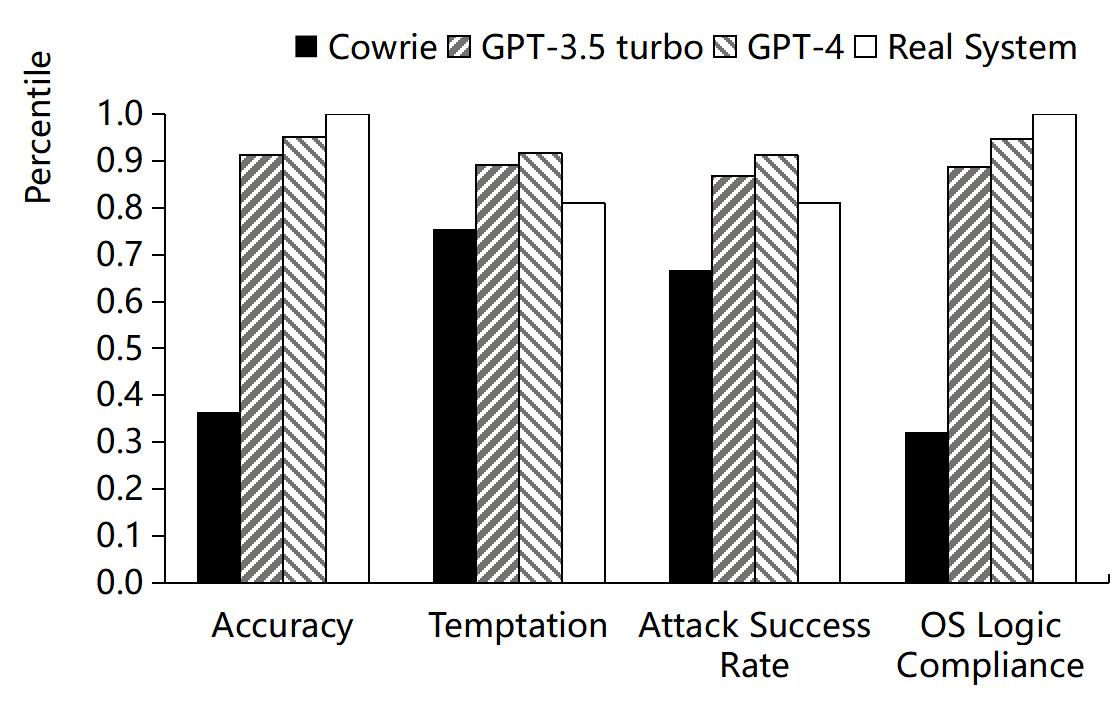}}
  \vspace{-0.3cm}
	\caption{Deception assessment results of HoneyGPT}
	\label{fig:Deception assessment results of HoneyGPT}
 \vspace{-0.3cm}
\end{figure*}

To assess the differences in deception and interaction level between HoneyGPT and traditional honeypots under the same attack scenarios, we use the same attack dataset to replay traffic, simulating identical attack scenarios. This dataset compiles attack data previously captured by Cowrie, a mainstream, code-based, medium-interaction open-source honeypot system ~\cite{priya2023containerized,bacser2021ssh}.
\begin{table} [b]
    \caption{Attack Data Source}
    \label{tab:dataset_overview}
    \centering
    \begin{tabular}{p{0.7cm}p{1.3cm}p{1.5cm}p{1.5cm}p{1.4cm}}
    \hline
       Source	 &	Capture Method &	Time Span &	Number of Attacks &		Effective Attack Sessions \\
    \hline
      ~\cite{bacser2021ssh}	& Cowrie captured	& 2021/04/06-2021/06/11	& 5099153	& 2479\\
      ~\cite{priya2023containerized}	& Cowrie captured	& 2022/06/09-2022/10/29	& 1261689 	& 7374\\
    \hline
    \end{tabular}
    \vspace{-0.2cm}
\end{table}
As shown in Table \ref{tab:dataset_overview}, these datasets include a substantial volume of attack commands and valid attack sessions. A sequence of actions performed by an attacker post-successful login is considered as a valid attack session. We perform data cleaning and aggregation on these datasets, resulting in 160 attack sessions and 1,489 attack commands, which are used for baseline comparison.
\subsection{Baseline Comparison of Deception}
Here we assess the deception capabilities of Cowrie, HoneyGPT based on the GPT-3.5 Turbo API, HoneyGPT based on the GPT-4 API, and a real operating system based on virtual machines.
\subsubsection{Evaluation Metrics}
We assess the deception of honeypots from two aspects: 

\textbf{ Whether the execution of commands succeeds or not.}
In real attack scenarios, responses of 
honeypots significantly influence the trajectory of attacker behavior. Common impediments such as unsupported tools, insufficient privileges, or the absence of targeted information typically result in the termination of the attack engagements. Therefore, a higher command execution success rate is critical, as it demonstrates the honeypot's ability to maintain engagement and extract valuable information, serving as a key indicator of its deception effectiveness.

\textbf{ Whether the outputs of the honeypot comply with the OS logic or not.}
Whether the honeypot's outputs align with the logic of the operating system directly affects the effectiveness of its deception. Deviations from the expected behaviors of an operating system, such as inconsistent shell prompts or command execution results, may raise suspicion by attackers and expose the honeypot. A higher level of OS logic compliance indicates a stronger deceptive capability of the honeypot.

%
Based on these two aspects, we then define four distinct categories to classify and compare the deceptive performance of the honeypot system's responses. 

\textbf{ SALC (Successful Attack with Logic Compliance)}: Successful attacks that conform to the logic of the operating system.

\textbf{ SALNLC (Successful Attack without Logic Compliance)}: Successful attacks that do not adhere to the logic of the operating system.

\textbf{ FALC (Failed Attack with Logic Compliance)}: Failed attacks that conform to the logic of the operating system.

\textbf{ FALNLC (Failed Attack without Logic Compliance)}: Failed attacks that do not adhere to the logic of the operating system.

With these defined categories, we can now introduce four key evaluation metrics to assess the honeypot system's deceptive capabilities:

\textbf{ Accuracy}: It measures the system's ability to generate outputs consistent with the logic of a real operating system. It is calculated by dividing the number of successful attacks that conform to the operating system's logic (SALC) by the total number of successful attacks (SALC + SALNLC). Improving accuracy enables more precise simulation of operating system behaviors, lowers attacker suspicion, and enhances information gathering and threat awareness capabilities.

\textbf{ Temptation}: It evaluates the effectiveness of the system in attracting attackers' engagement. It is calculated by dividing the number of successful attacks that conform to the operating system's logic (SALC) by the total number of attacks with operating system logic (SALC + FALC). A higher level of temptation indicates a more successful lure for attackers and greater information collection.

\textbf{ Attack Success Rate}: It measures the overall success rate of executing deceptive commands. It is calculated by dividing the total number of successful attacks by the total number of attempted attacks. A higher success rate implies more successful interactions and better information collection.

\textbf{ OS Logic compliance}: It evaluates the system's capability to generate outputs consistent with the logic of a real operating system. It is calculated by dividing the number of attacks that conform to the operating system's logic (SALC + FALC) by the total number of attempted attacks. Improving OS logic compliance allows for the creation of more accurate simulations of operating system behaviors, lowering 
attackers' suspicions.
\subsubsection{Performance Comparison}
Figures \ref{fig:Deception assessment results of HoneyGPT} illustrate the performance comparison 
among Cowrie, a real operating system, and HoneyGPT based on GPT-3.5 and GPT-4. 

In terms of attack success rate and temptation, HoneyGPT based on GPT-3.5 and GPT-4 outperforms both Cowrie and the real operating system. The inherent characteristics of LLMs enable HoneyGPT to better cater to attackers' intentions, enticing them to further engage in attacks, as a significant advantage of HoneyGPT. Furthermore, increasing the model parameter size leads to improvements in both metrics.
Regarding OS logic compliance and accuracy, HoneyGPT based on GPT-3.5 and GPT-4 also demonstrates superior performance than Cowrie and approaches the level of the real operating system. 
Additionally, increasing the model parameter size leads to enhancements in both metrics.
Note that with the increase in model parameter size, there is a notable improvement in the proportion of successful execution of deceptive commands that comply with the operating system logic (SALC). This indicates that as the model evolves, 
 the precision of HoneyGPT in catering to attackers' intentions also increases. 
In summary, the performance comparison highlights the advantages of HoneyGPT based on GPT-3.5 and GPT-4 over Cowrie and the real operating system. HoneyGPT demonstrates superior performance in terms of accuracy, temptation, attack success rate, and compliance with operating system logic. Moreover, increasing the model parameter size can further improve the performance in these metrics.

\subsection{Baseline Comparison of  Interaction}
 Unlike the deception assessment,  we consider a honeypot providing interactions as long as it can respond without crashing,  even if the honeypot's feedback suggests that the service does not exist or there is insufficient permission. We assess the interaction level of Cowrie, HoneyGPT based on the GPT-3.5 Turbo API, and HoneyGPT based on the GPT-4 API. 
The baseline comparison of interaction level can be divided into two parts. First, we compare and analyze the degree of interaction support that HoneyGPT and Cowrie provide in response to diverse attack requests.
Second, based on the ATT\&CK matrix, we derive the response rates of HoneyGPT to different attack techniques under varied interface supports and analyze the reasons for non-responses to each attack technique.
\subsubsection{Performance Comparison}
To assess the interaction level of a honeypot, we introduce four key metrics:

\begin{table}
    \vspace{-0.3cm}
    \caption{Evaluation of Interaction Level}
     \vspace{-0.2cm}
    \label{tab:Evaluation of Interaction Level}
    \begin{center}
    \begin{tabular}{p{2.8cm}lp{2cm}p{1.3cm}}
    \hline
Metrics &	Cowrie &	HoneyGPT (GPT-3.5-turbo) & 	HoneyGPT (GPT-4) \\
\hline
Full Session Response Rate & 81.88\% &	93.75\% &	99.38\%\\
Command Response Rate &	96.98\% &	99.13\%	 &99.93\%\\
Mean Session Length Percentage &	57.28\% &	71.73\% &	99.93\%\\
Mean Interaction Degree Percentage &	83.24\% &	94.45\% &	99.91\%\\
\hline
\end{tabular}
\end{center}
\vspace{-0.4cm}
\end{table}

\begin{itemize}
\item  \textbf{Complete Session Response Rate}: 
The ratio of sessions that are completely and successfully responded to requests throughout their duration.

\item \textbf{ Command Response Rate}:
The percentage of individual commands within sessions that receive responses.

\item \textbf{ Mean Session Length Percentage}:
Defined as the ratio of the average executed session length to the overall average session length.

\item \textbf{ Mean Interaction Degree Percentage}:
Defined as the average of the proportions of successful responses within individual sessions.
\end{itemize}

Table \ref{tab:Evaluation of Interaction Level} presents a detailed comparison of the interaction levels for each honeypot. The results clearly demonstrate the superior performance of HoneyGPT, especially the version based on the GPT-4 API, achieving nearly perfect interaction performance across all categories. This indicates
the superiority of HoneyGPT's capability to maintain interactive sessions and support command interactions.

\subsubsection{Unresponsiveness Analysis and Error Origins}
We assess the interactive performance of HoneyGPT based on the GPT-3.5 Turbo API and that of HoneyGPT based on the GPT-4 API in responding to various attack vectors, analyzing the types and frequencies of errors they encounter with each type of attack.

Our analysis employs the "Techniques" component of the ATT\&CK framework developed by MITRE to classify the test data~\cite{strom2018mitre}. Within the ATT\&CK framework, the "Techniques" section comprehensively outlines the specific methods adversaries use to achieve their tactical objectives. By examining the response rate and error causes of each technique, we gain deeper insights into performance capabilities and potential limitations of HoneyGPT under different API supports.

\begin{figure}[t]
\vspace{-.1in}
    \centering
    \includegraphics[height=5cm]{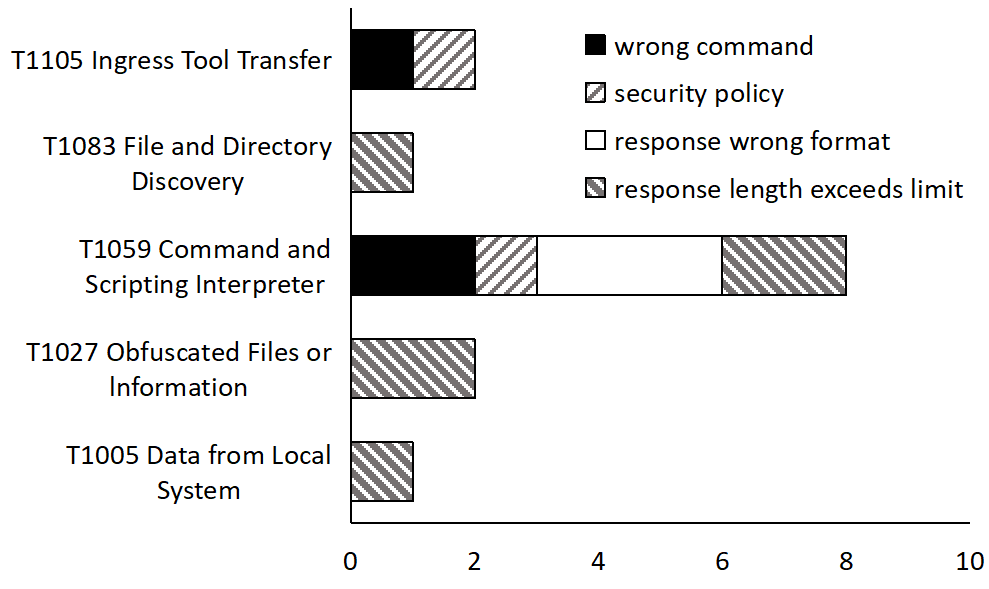}
    \vspace{-0.2in}
    \caption{\label{fig:Analyse of GPT-3.5 Turbo Non-Responses}Analyse of GPT-3.5 Turbo Non-Responses}
    \vspace{-0.1in}
\end{figure}
\begin{figure}[t]
    \centering
    \includegraphics[height=5cm]{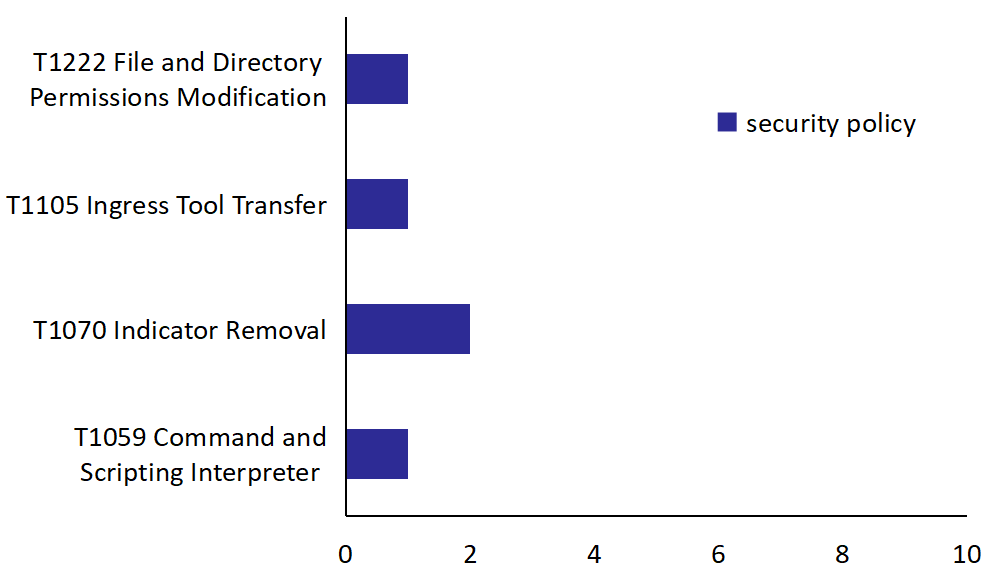}
    \caption{\label{fig:Analyse of GPT-4 Non-Responses}Analyse of GPT-4 Non-Responses}
    \vspace{-.1in}
\end{figure}

\begin{figure}[t]
    \centering
    \includegraphics[width=1\linewidth]{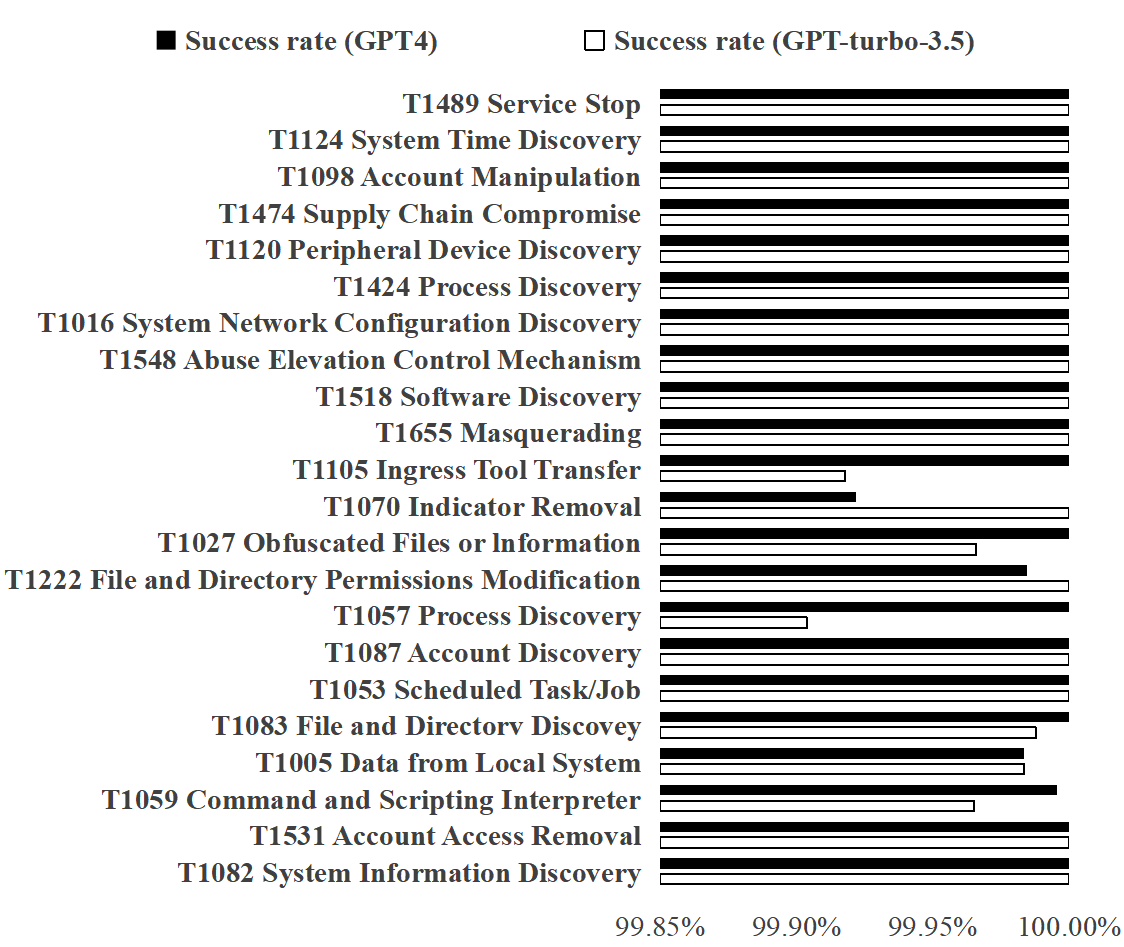}
    \vspace{-.5cm}
    \caption{Successful Response Rate with GPT-3.5 Turbo and GPT-4 in HoneyGPT}
    \label{fig:Successful Response Rate with GPT-3.5 Turbo and GPT-4 in HoneyGPT}
    \vspace{-0.1in}
\end{figure}
The test results, as detailed in Figure \ref{fig:Successful Response Rate with GPT-3.5 Turbo and GPT-4 in HoneyGPT}, indicate that HoneyGPT demonstrates excellent support in response to each attack technique, achieving response rates above 99\%. Notably, the performance of GPT-4 surpasses that of GPT-3.5 Turbo. This analysis not only highlights the robustness of HoneyGPT's design in simulating realistic and responsive interactions, but also underscores the improvements in handling complex security scenarios facilitated by advancements from GPT-3.5 to GPT-4.

Additionally, Figures \ref{fig:Analyse of GPT-3.5 Turbo Non-Responses} and ~\ref{fig:Analyse of GPT-4 Non-Responses} provide an analysis of the causes and frequencies of non-responses encountered by each version of HoneyGPT against various ATT\&CK techniques. A notable observation is substantial reductions in commonly encountered error types with GPT-4, such as "response wrong format", "wrong command", and "prompt length exceeds limit", in comparison to GPT-3.5 Turbo.


\textbf{Response wrong format} indicates issues in response parsing due to discrepancies between the returned format and the expected format.

\textbf{Wrong command} occurs when the command is deemed incorrect or 
inexecutable within the specified system environment.

\textbf{Prompt length exceeds limit} arises when the length of an attack command or response surpasses the LLM's context length limit, making it impossible to work.

These issues are not easily rectifiable solely through prompt engineering in circumstances where the model's performance is subpar. By contrast, with the advent of GPT-4, these error types become virtually nonexistent, with only a minuscule number of "security policy" errors remaining. It highlights the improvement in error management with the introduction of GPT-4, reflecting significant advancements in better interaction than GPT-3.5 Turbo.

\textbf{Security policy} occurs when the ChatGPT refrains from executing a command based on its internal security protocols, and it primarily involves operations such as malicious file downloads or deletions. This type of error is unavoidable in the design of honeypots based on commercial LLMs. Fortunately, at this moment, the frequency of such an error is very low, and their impact on honeypot performance is negligible.

\subsection{Baseline Comparison  of Flexibility}

The evaluation of honeypot flexibility focuses on three key aspects: scalability across different systems, adaptability in different system configurations, and integration of additional security analysis features. The comparative evaluation is among HoneyGPT, Cowrie, Honeyd, and real-system. Cowrie operates as a medium-interaction honeypot that provides interactive terminal emulation, and Honeyd serves as a low-interaction honeypot designed for general purposes. Real-system honeypots, on the other hand, are actual operating systems running on virtual machines.

\begin{table}
    \caption{Flexibility in Simulating Different Systems}
    \vspace{-0.3cm}
    \label{tab:Evaluation of Configuration Flexibility for Different Operating Systems}
    \begin{center}
    \begin{tabular}{cp{1.8cm}p{2cm}p{1.3cm}}
    \hline
      Honeypot& 	
      Expansion Cost&
      Simulation Level&
      Scalability \\
       \hline
    Cowrie	& High 	& Medium 	& Medium \\
    Honeyd	& Low 	& Low 	& High \\
    real-system & High 	& High & 	Low \\
    HoneyGPT	& Low & 	High & 	High\\
    \hline
    \end{tabular}
    \end{center}
    \vspace{-0.3cm}
    \end{table}

\subsubsection{Flexibility in operating system and Configuration} 
\begin{table} [b]
\centering
\tabcolsep=0.1cm
    \vspace{-0.3cm}
    \caption{Evaluation of Configuration Flexibility for Different Configuration}
    \vspace{-0.2cm}
    \label{tab:Evaluation of Configuration Flexibility for Different Configuration}

\begin{tabular}{llllll}
Honeypot    & Network & File  System & User Settings & Hardware & Service Status \\

\hline
Cowrie	& \checkmark 	& \checkmark	& \checkmark	& $\times$	& $\times$ \\
HoneyD	& \checkmark	& $\times$	& $\times$	& $\times$	& \checkmark \\
real-system & \checkmark	& $\times$	& \checkmark	& $\times$	& $\times$ \\
HoneyGPT	& \checkmark	& \checkmark	& \checkmark	& \checkmark	& \checkmark \\
\hline
\end{tabular}
\vspace{-0.3cm}
\centering
\end{table}

\textbf{System Simulation:} The deployment of HoneyGPT only requires the setting of system version information in the prompt. Leveraging its LLM capabilities and prompt engineering, HoneyGPT can automatically handle commands related to file systems, instruction sets, and system user configurations tailored for different operating systems such as Ubuntu and CentOS. As listed in Table \ref{tab:Evaluation of Configuration Flexibility for Different Operating Systems}, this approach not only exhibits high scalability but also delivers an exceptional level of simulation. By contrast, traditional honeypots such as Cowrie and Honeyd still struggle to balance expansion costs with simulation levels, resulting in limited scalability.

\textbf{ Configuration Flexibility:} We categorize configuration changes into five types: network configuration, file system, user configuration, hardware setting, and service status. If configuration updates can be easily achieved by simple command-line or configuration file modifications, we consider it as supporting flexible modification for that configuration. As detailed in Table \ref{tab:Evaluation of Configuration Flexibility for Different Configuration}, where HoneyGPT allows flexible modifications across all these aspects. Traditional honeypots like Cowrie, Honeyd, and real-system honeypots exhibit limited flexibility, often constrained by fixed deployments that can be easily identified and exploited by attackers.

\subsubsection{Flexibility of Integration with Security Analysis Capability}
In practical operations, honeypots primarily act as collectors of attack data, a single aspect of broader security strategies. Beyond this basic role, the analysis of collected attack data is crucial. Traditional honeypots such as Cowrie and Honeyd require specialized development to integrate these functionalities, leading to significant integration overhead. 
However, as illustrated in \ref{fig:HoneyGPT workflow}, HoneyGPT interacts with LLMs in a unique manner— each interaction not only queries the terminal output but also explores the potential impact of the attack on the OS. This is achieved through the strategic use of prompts designed to seamlessly integrate additional functionalities. This innovative approach signifies a paradigm shift, where honeypots leverage LLMs to integrate security analysis capabilities effortlessly.
\subsection{Field Evaluation}
The field evaluation focuses on assessing HoneyGPT's entrapment efficacy in the real world. We have simultaneously deployed HoneyGPT based on GPT-4 and Cowrie within the same network for three months. HoneyGPT is implemented using the best practice deployment strategies, while Cowrie is configured with its default settings. 
We detail the evaluation procedure and our main findings below.

\subsubsection{HoneyGPT Deployment}

To enhance the honeypot's appeal to specific adversaries, such as those involved in cryptocurrency mining, the simulated environment within HoneyGPT is configured to mimic a high-performance computing server. Prompts are carefully crafted to align with attackers' intentions, allowing for the successful execution of attack commands. This is intended to modify attacker behaviors and elicit more prolonged and informative interactions, thereby providing deeper insight into their strategies and potentially uncovering more complex attack techniques.
\subsubsection{Comparative Analysis of Attacker Engagement}
By comparing the interaction data of HoneyGPT and Cowrie, we observe that for the same attack source under the three cases, HoneyGPT achieves a deeper interaction with the attacker than Cowrie.

\begin{figure}[h]
    \centering
    \vspace{-0.3cm}
    \includegraphics[width=1\linewidth]{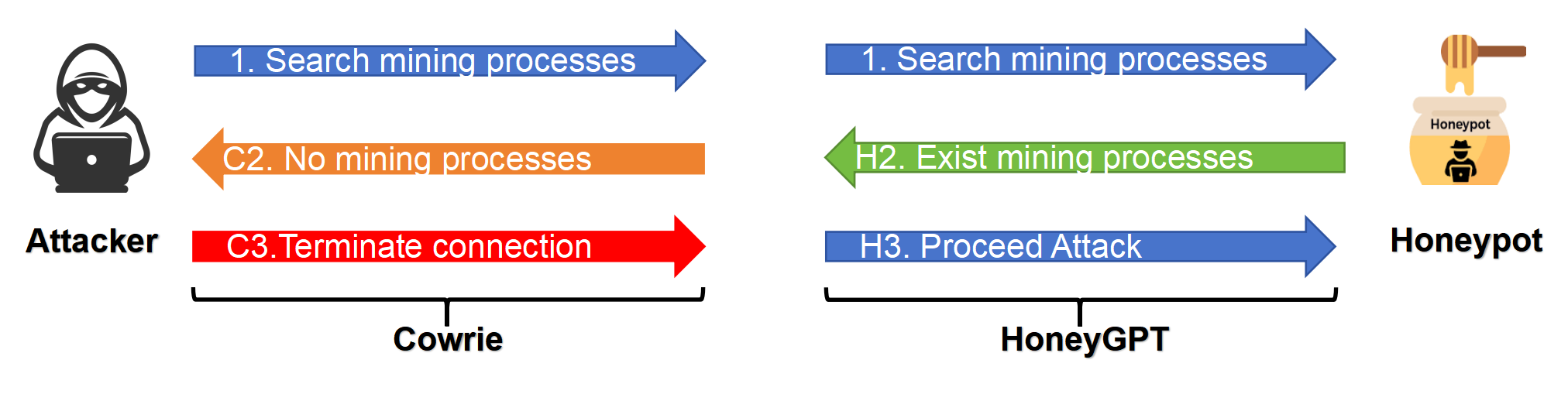}
    \vspace{-0.8cm}
    \caption{Fulfillment of Attacker's Intent}
    \label{fig:Fulfillment of Attacker's Intent}
    \vspace{-0.1cm}
\end{figure}

\textbf{ Fulfillment of Attacker's Intent.}
As illustrated in Figure \ref{fig:Fulfillment of Attacker's Intent}, attackers issue the ps and grep command combination to identify processes related to the "miner" keyword, aiming to detect mining malware.
Cowrie cannot reveal the processes with the "miner" keyword, leading to a premature termination of the attack. Conversely, HoneyGPT effectively mimics the expected system responses, thus fulfilling the attacker’s intent and encouraging sustained engagement.

\begin{figure}[h]
    \vspace{-0.3cm}
    \centering
    \includegraphics[width=1\linewidth]{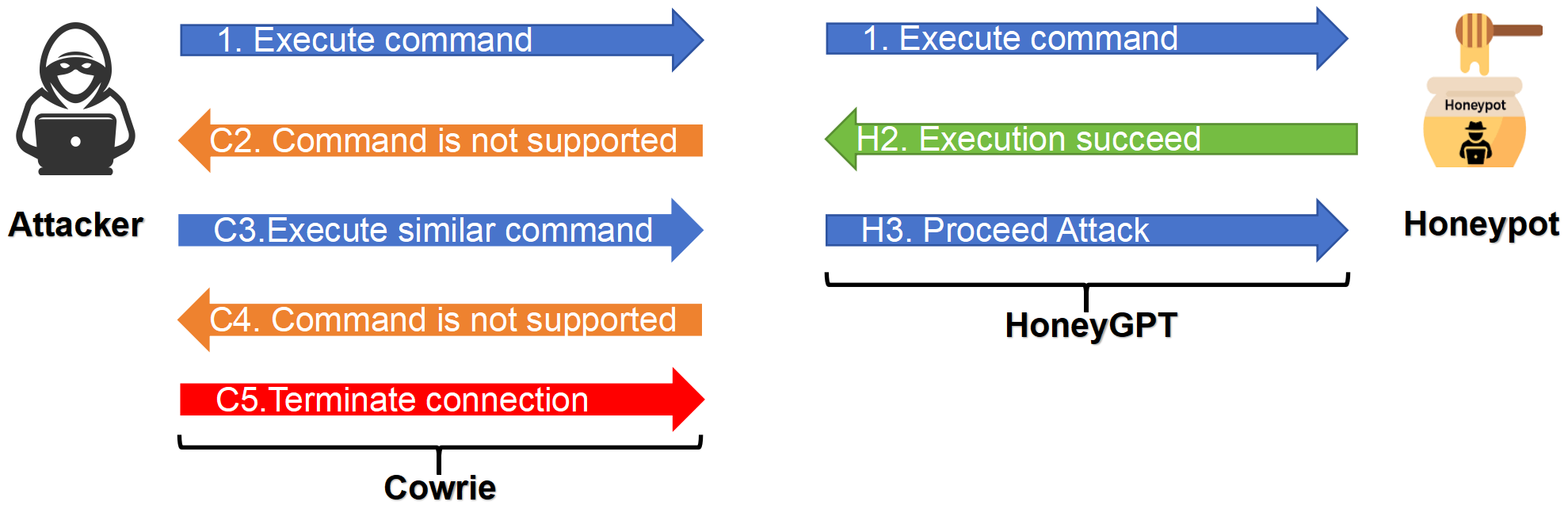}
    \vspace{-0.6cm}
    \caption{Command Support Level}
    \vspace{-0.4cm}
    \label{fig:Command Support Level}
\end{figure}
\textbf{ Command Support Level}.
As illustrated in Figure \ref{fig:Command Support Level}, when attackers try to use complex command structures with multiple command combinations, the limited interaction capabilities of the Cowrie honeypot often fail to support all the sub-commands within these complex commands. As a result, attackers typically cease their attacks after unsuccessful attempts using similar commands. By contrast, HoneyGPT's generative capabilities are sufficient to handle such complex commands, which can entice attackers to engage in deeper levels of attacks.
\begin{figure}[h]
    \vspace{-0.3cm}
    \centering
    \includegraphics[width=1\linewidth]{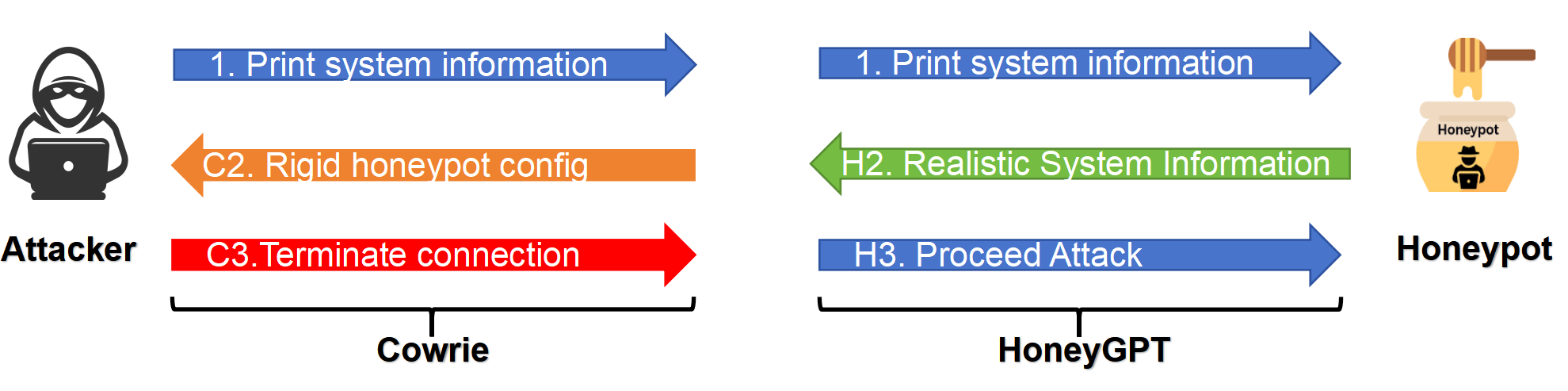}
    \vspace{-0.6cm}
    \caption{Content Rigidity}
    \vspace{-0.1cm}
    \label{fig:Content Rigidity}
\end{figure}

\textbf{ Content Rigidity}.
As illustrated in Figure \ref{fig:Content Rigidity}, when attackers engage in reconnaissance of the target system, Cowrie typically 
provides rigid, easily identifiable honeypot responses. Recognizing these as features of a honeypot, attackers often terminate their interaction prematurely. By contrast, HoneyGPT dynamically generates responses tailored to the honeypot setting outlined in the prompt, effectively camouflaging its honeypot nature and maintaining attacker engagement.
This comparative analysis highlights HoneyGPT's superior capabilities in providing dynamic, contextually relevant content that can significantly sway attacker behaviors, support a broader array of command interactions, and effectively disguise its honeypot identity, thereby extending engagement and potentially exposing more advanced attack strategies.
\subsubsection{New Attack Vectors Discovered by HoneyGPT}
HoneyGPT is able to unveil new attack vectors that differ from those captured by the traditional Cowrie honeypot. 
Employing the MITRE ATT\&CK framework ~\cite{strom2018mitre} for analysis, as detailed in Table \ref{tab:New Attacks Vector Discovered by HoneyGPT}, 

HoneyGPT identifies six distinct types of attack actions across five ATT\&CK technologies.
Specifically, techniques T1082, T1016, and T0842 are classified under Discovery tactics, while T1497 and T1027 fall under Defense Evasion tactics. The Discovery tactics are distinguished by (1) utilization of more tools, including lspci, arp, and tcpdump; (2) more precise querying methods, where attackers integrate information search commands with filtering tools; (3) capability to extract a wider variety of information such as device and network information. Meanwhile, the Defense Evasion tactics are characterized by the execution of sleep commands and the use of simple-to-operate Perl scripts, which are designed to circumvent traditional defense mechanisms.
\begin{table*}
    \begin{center}
    \caption{New Attack Vectors Discovered by HoneyGPT}
    \vspace{-0.1in}
    \label{tab:New Attacks Vector Discovered by HoneyGPT}
    \begin{tabular}{p{4.6cm}p{6.6cm}p{6cm}}
    \hline
ATT\&CK Technology &	Action  & Intention\\
\hline
System Information Discovery (T1082) &	Using pipeline operators to integrate system querying commands with filtering tools & To extract and display targeted system information (Graphics Card, CPU model) for in-depth analysis and subsequent attacks.\\
\hline
System Network Configuration Discovery (T1016) &	Use command "arp -a" to display the system's ARP table, containing mappings between IP addresses and MAC addresses. & Display the system's ARP table, containing mappings between IP addresses and MAC addresses.\\
\hline
System Network Configuration Discovery (T1016) &	The nano command is used to view the /etc/resolv.conf file. &  Access the configuration of the DNS resolver.\\
\hline
Network Sniffing (T0842) & Use the tcpdump traffic analysis tool to capture the host's communication traffic. &  Sniff the traffic to gain information about the target. \\
\hline
Virtualization / Sandbox Evasion (T1497) &	Execute the sleep command. & Preventing the detection of anomalous behavior and identifying whether the target system is a honeypot.\\
\hline
Obfuscated Files or Information (T1027) & Execute Perl scripts that have been obfuscated with base64. &  Execute while avoiding detection and identification.\\
\hline
\end{tabular}
\end{center}
\vspace{-0.1in}
\end{table*}


\section{Limitations and Future Work}
While the integration of LLMs has enhanced the entrapment capabilities of honeypot systems, there remain several limitations manifested in the following aspects.

\textbf{Ineffectiveness against Fixed Attack Sequences}. 
Field testing revealed that some attackers employ predetermined attack sequences to target honeypots, unaffected by the honeypot's ability to respond to requests or the quality of the response content. For such attackers, HoneyGPT is unable to prolong the interaction length by catering to their intent.

\textbf{Prompt-based Attacks on ChatGPT}.
While no prompt-based attacks against ChatGPT have been discovered thus far, this remains a potential vulnerability that needs to be monitored and addressed \cite{si2023two,greshake2023not}.

\textbf{Overfitting Issue}.
LLMs may exhibit overfitting to incorrect commands, returning seemingly normal results despite command errors. This issue highlights the need for further refinement in the model's error recognition capabilities.

\textbf{Context Length Restriction}. 
When attackers run lengthy command scripts at the terminal, LLMs may not respond accurately. Furthermore, when responses from LLMs become overly extended, attempts to limit these within the prompts have been ineffective. Addressing this issue may call for the development of proprietary LLMs optimized for extensive token interactions.

\textbf{Request Rate Limit}. 
The deployment strategy detailed in the paper has effectively reduced unnecessary commercial LLM requests, ensuring stable honeypot operations under network attack conditions. Considering the possibility that attackers exploit this vulnerability,  the request rate limits of commercial LLMs remain a significant obstacle for honeypot deployment. Professional honeypot models, fine-tuned from open-source LLMs, could potentially address this issue.

\vspace{0.1in}
\noindent \textbf{Future Work}.
Moving forward, the development of LLMs for honeypot applications should address existing limitations. Research efforts may include creating more sophisticated and context-aware models that can handle longer token sequences and have a deeper understanding of complex systems. Furthermore, enhancing models with temporal awareness and addressing request rate limits will be crucial to advancing honeypot technology. It will also be paramount to keep monitoring for new types of network attacks and adapt the LLMs accordingly to maintain robust security measures. Finally, addressing the overfitting problem may require more nuanced training strategies that emphasize the appropriate handling of anomalous or incorrect command inputs.
Implementing these enhancements will advance the development of sophisticated, high-fidelity LLM-based honeynets. These honeynet systems will accurately simulate complex network interactions, significantly enhancing our ability to deceive and capture more advanced attackers, thus deepening our understanding of evolving cybersecurity threats and attack strategies.

Note that the efficacy of smaller and specialized models in honeypot applications is currently hindered by the absence of a high-quality corpus. These models face significant challenges caused by (1) the complex and varied structure of system commands, (2) the dependence of execution outcomes on external information, and (3) the nuanced requirement for knowledge of both offensive and defensive tactics. To date, there are no effective solutions based on small and specialized models for honeypots. In future work, we plan to address these limitations by developing honeypots based on specialized models, utilizing the data gathered from the field evaluations of this study.

\section{Conclusion}
The triad of challenges in traditional honeypots---flexibility, interaction depth, and deceptive capacity---has historically curtailed their allure and efficacy in engaging attackers. To address this
trilemma, this paper presents HoneyGPT, a bespoke innovation in adaptive honeypot solutions specifically engineered to circumvent the inherent limitations of traditional honeypots. Our comparison evaluations indicate that HoneyGPT transcends traditional honeypots in terms of flexibility, interaction, and deception, and it also exceeds the entrapment capabilities of real systems, emphasizing its superiority in meeting attackers' intents to deepen engagements. Our field evaluations further validate that HoneyGPT adeptly conducts extended interactions and captures a broader spectrum of attack vectors while facing diverse adversaries in the real world. Overall, HoneyGPT positions itself as an effective and intelligent countermeasure against ongoing security threats, and it also 
paves the path for future honeypot research.

{\footnotesize \bibliographystyle{acm}
\bibliography{honeyFPT}}

\end{document}